\documentclass[aps,showpacs,preprintnumbers,amsmath,amssymb�eqsecnum,
twocolumn, tightenlines,
]{revtex4}

\usepackage[dvips]{graphicx}
\usepackage{bm}
\usepackage{longtable}
\usepackage{amsmath}
\usepackage{dcolumn}
\usepackage{placeins}
\usepackage{expdlist}

\begin{document}

\bibliographystyle{apsrev}

\title {Neutrino Velocity and the Variability of Fundamental Constants}
\author{ 
V. V. Flambaum$^{\,(a)}$ and M. Pospelov$^{\,(b,c)}$} 
\affiliation{
$^{\,(a)}$ School of Physics, University of New South Wales, Sydney
  2052, Australia\\
$^{\,(b)}${\it Department of Physics and Astronomy, University of Victoria, 
     Victoria, BC V8P 1A1, Canada}\\
$^{\,(c)}${\it Perimeter Institute for Theoretical Physics, Waterloo,
ON N2J 2W9, Canada}}

    \date{\today}

    \begin{abstract}
 

Neutrino speed experiments could be viewed not only as tests of Lorentz invariance but also as  
measurements of limiting propagation speed for all 
standard model species below certain depth where no direct 
metrological information is available. 
The latter option, hypothetically caused by some 
chameleon-type background, could be tested in the next installment of the neutrino speed experiments.
We also show that that complementary constraints on the same class of models can be obtained
 with  experiments testing clock 
universality in deep underground/underwater experiments. 
By considering the explicit QED model with particle-universal 
modification of propagation speed by a depth-dependent tensor background, we show that 
in general one should expect larger-than-GR shifts of the clock frequencies and clock non-universality.
This can be tested by comparison of the narrow transitions in atomic clocks, which for a generic model 
could deliver a superior accuracy compared to the neutrino speed experiments.

    \end{abstract}

\pacs{
              04.50.Kd
                }

\maketitle

\section{Introduction}
  The OPERA experiment \cite{Opera} created a stir of theoretical activity 
by suggesting that the speed of muon neutrino may
 be higher than the speed of light in vacuum,
\begin{eqnarray}
\epsilon_{\rm OPERA} \equiv \frac{c_{\nu}-c_{\gamma,{\rm vac}}}{c} =(2.37\pm 0.32^{+0.34}_{-0.24})\times 10^{-5},
\label{opera}
\end{eqnarray}
with the significance of $\sim 6 \sigma$. 
These results agree with earlier measurement by the MINOS collaboration \cite{Minos}
that also found $\epsilon$ to be positive, consistent with (\ref{opera}), and yet consistent with zero at $2\sigma$,
 \begin{eqnarray}
\epsilon_{\rm MINOS}  =(5.1\pm 2.9)\times 10^{-5}.
\label{minos}
\end{eqnarray}
While the claimed OPERA measurement has been shown to result from the initially undetected instrumental 
error, 
a new measurement by ICARUS \cite{Icarus} is in perfect agreement with standard physics expectations,
\begin{eqnarray}
\epsilon_{\rm ICARUS}  =(0.01\pm 0.16\pm  0.37)\times 10^{-5}.
\label{icarus}
\end{eqnarray}
The initial results from OPERA  led to the number of experiments testing the neutrino 
propagation speed, and more results are expected within the short time frame. Given 
 that the neutrino speed experiments are reaching the 10 ppm (or better) 
 accuracy of their measurements of $c_\nu$, we ask the question whether such tests have additional 
physics motivations, given that the original result (\ref{opera}) is incorrect.

In this paper, we shall argue that the neutrino speed measurement provides an interesting probe 
of fundamental constants at depths that are not yet accessible to direct metrological experiments. 
While we do not explicitly construct a dynamical model that could give $\epsilon \sim 10^{-5}$, 
we settle for an intermediate framework 
with the universal density-dependent modification of the limiting propagation speed for all 
matter species. The purpose of this note is to point out that  additional 
experiments with ordinary atoms  could also  test deviations of $c$ from $c_{\gamma, \rm vac}$ deep underground/underwater.

Would continuation of the neutrino speed experiments provide nontrivial probes of Lorentz invariance? 
Despite last months numerous attempts to construct models that could "fit" the OPERA 
result, we note that none of the models with density-independent global modification of Lorentz invariance 
for neutrinos seem to work. Below we summarize the 
main arguments why large Lorentz violation (LV) for neutrinos is highly unlikely:
\begin{enumerate}

\item As pointed out by Cohen and Glashow \cite{CG}, the neutrino propagation faster than electron's 
limiting speed at the level of (\ref{opera})  would result in rapid energy loss by neutrinos,
which contradicts observations of energetic atmospheric and beam neutrinos. Moreover, if speed 
of neutrinos is higher than the speed of quarks, a pion would not decay neutrinos  above 
certain energies \cite{nopiondecay}. Note that the Cohen-Glashow argument applies not only to models 
that postulate global LV for neutrinos, but also to models that exploit possible local modification of 
neutrino speed compared to other species caused {\em e.g.} by a tensor background \cite{DV}.
(For a divergent point of view see, however, Ref. \cite{Brod}.)

\item Large global violation of Lorentz invariance for neutrinos is in conflict with apparently 
normal timing of their arrival from the explosion of 1987a supernova.

\item Large global violation of Lorentz invariance for neutrinos is difficult to reconcile 
with tightly constrained LV parameters for electrons, as the two sectors are connected by the 
$W$-boson loop \cite{MP,Sergei}.

\end{enumerate}

On account of these constraints, it is highly unlikely that Lorentz invariance is broken for neutrinos 
either globally or locally at such a large level as $\epsilon \sim 10^{-5}$. 
Therefore, the direct measurement of $c_\nu$ does not provide a superior 
test of Lorentz invariance for neutrinos. On the other hand, it is possible to speculate \cite{Heb,Brax} that the limiting propagation speed is somehow modified {\em for all species} in the Earth's interior, 
and that neutrino speed experiments are the tests of this possibility.  
The  neutrinos in OPERA and ICARUS experiments propagate 
at average depth of $\sim6$ km, reaching $\sim 11$ km in the middle. There are no direct tests 
of propagation speed performed at such depths, and therefore a possibility of the common to all particle 
change in $c$, however far-fetched, is not excluded. Therefore, the results (\ref{opera}-\ref{icarus}) can be 
considered as tests of such possibility. 
One should also recall that gravitational field itself modifies the propagation speed 
for all species, but the modifications suggested by (\ref{opera}) are many orders of magnitude larger.
If indeed a propagation speed is modified at some depth below Earth's surface,  one acquires two additional requirements for a model of this type: 
\begin{description}
\item  4. In-medium modification should sharply decrease near the Earth's surface in 
order to be in accord with precise measurements of the gravitational force and the limiting propagation 
speed.
\item 5. Whatever backgrounds modify the propagation speed in the Earth's interior, they
should not couple to the $00$ component of the stress energy for matter fields in order
to avoid larger-than-gravity forces inside planets/stars. 
\end{description}

Models where in-medium properties of Lorentz invariant physical parameters such as 
masses and coupling constants are different from the same values in vacuum were introduced 
a few years ago \cite{OP}. They are related to previous ideas about the modification of 
the scalar-induced gravitational force by the presence of matter overdensities \cite{DN,PD,KW}. 
(Density dependence rather than redshift dependence could be an alternative interpretation 
of the non-zero result of Ref. \cite{Webb} that looked for the variation of $\alpha_{EM}$ in absorption 
systems at cosmological distances.) Admittedly, models of 
in-medium modifications of the propagation speed are  harder to construct, as they would require 
"condensation" of fields with non-trivial Lorentz indices. 

Assuming for a moment that some variants of density-dependence could modify the limiting propagation 
speed of neutrinos and all other species, we ask the question whether additional measurements performed 
with ordinary matter (not neutrinos) could test $c$ and other constants deep underground/underwater. 
In the next section we discuss to what extent the (depth-induced) variation of the limiting 
propagation speed can lead to the variation of coupling 
constants and clock non-universality, and in the concluding section 
we propose new experiments that could detect such effects. 

\section{Limiting propagation speed and changing couplings}

The propagation speed of any matter species 
can be modified by the non-Lorentz invariant backgrounds. Consider a 
scalar field, with the
Lagrangian modified by some tensor background $h_{\mu\nu}$:
\begin{equation}
\label{onefield}
{\cal L}_\phi = \frac12 (\partial_\mu \phi)^2 - \frac12 m^2\phi^2  
+ h_{\mu\nu} \partial_\mu \phi \partial_\nu \phi.
\end{equation}
Here we do not distinguish between upper and lower indices and perform the
Lorentz summations with $\eta_{\mu\nu} = {\rm diag}(1,-1,-1,-1)$. 
Moreover, we explicitly work in the system of units where $\hbar=c=1$, and the limiting velocity 
is $1$ if the tensor background $h_{\mu\nu}$ vanishes.
The limiting velocity for 
the $\phi$ particle travelling along $z$ direction ($E\gg m$) is given by 
\begin{equation}
c^2_\phi = \frac{1-2h_{zz}}{1+2h_{00}}\simeq 1 - 2h_{00} -2h_{zz}.
\end{equation}
One can see that negative $h_{00}$ and/or  $h_{zz}$ can "speed-up" the $\phi$-field via effectively 
stretching time  or shortening distances. Besides the simplest possibility (\ref{onefield}), one 
could introduce higher dimensional operators, $h_{\mu\nu\lambda\kappa} \phi \partial_\mu\partial_\nu\partial_\lambda\partial_\kappa \phi$ that will lead to the energy-dependent 
modification of the propagation speed. 

For the rest of this paper we use the following ansatz for the $h_{\mu\nu}$ field
\begin{equation}
h_{00}=h_{0i}=0;~~~h_{ii} = -\epsilon\times{\rm diag}(1,1,1), 
\label{epsilon}
\end{equation}
so that (\ref{onefield}) can be 
written as ${\cal L}_\phi = \frac12(\partial_t \phi)^2 - \frac12(1+2\epsilon)(\partial_i\phi)^2$, 
and we introduce the same modification for all fields of the standard model. The spatial anisotropy of $\epsilon$ is not 
a crucial ingredient and is assumed for convenience. To comply with all requirements listed in the introduction, we take
$\epsilon\equiv \epsilon({\rm depth})$ to be some sharp function of the depth, with $\epsilon=0$ at the Earth's surface. 
The initial super-luminal claim by OPERA  in our model  would imply that 
at some critical depth $z_0$ one should assume that $\epsilon$ deviates from 
0 and develops a positive value. The exact relation between $\epsilon$ in (\ref{epsilon}) and 
$\epsilon_{\rm OPERA}$ would depend on $z_0$ in a simple geometric way, but regardless of that 
measurement (\ref{opera}) would imply $\epsilon\geq 2.5\times 10^{-5}$, while of course (\ref{icarus}) is 
perfectly consistent with $\epsilon = 0$.

One could attempt replacing the tensor background (introduced here by hand) with some dynamical scalar, vector, or tensor fields: $s$, $V_\mu$ or $H_{\mu\nu}$. 
Going over to the canonical normalization of the kinetic terms for these fields, one can 
write down  the interactions modifying the propagation speed of a generic SM field $\phi$, 
$M^{-4}(\partial_\mu s \partial_\mu \phi)^2; ~M^{-2} (V_\mu \partial_\mu \phi)^2;~M^{-1}H_{\mu\nu}\partial_\mu \phi \partial_\nu \phi$ \cite{DV}. The immediate problem with the first two constructions is that in order to have any connection with OPERA result, the scale $M$ would have to be exceedingly low. The most recent model-building attempt  to reconcile OPERA measurement with observations \cite{Brax}
 employs a scalar field, and $M$ has to be below an MeV. Unfortunately, this is in plain contradiction with 
direct particle physics experiments.  ({\em E.g.} electron-positron scattering  remains consistent with the prediction of 
the SM to energies of $\sim$100 GeV, while the double-$s$ exchange 1-loop diagram would lead to the significant modification 
of scattering for any $M$ below O(10 GeV).) The tensor background offers perhaps the only reasonable hope 
for the dynamical model \cite{DV}. Still, even if we leave aside theoretical issues 
with UV completion, it appears difficult if not impossible to construct a weakly coupled model where 
dynamical $H_{\mu\nu}(z)$ follows gravitational potential profile and does not run into contradiction with some
observations. The model of Ref. \cite{DV} that uses 
much enhanced coupling of $H_{\mu\nu}$ to neutrinos faces a problem of 
Cerenkov radiation of electron-positron pairs, and an attempt to cure it by postulating the same interaction 
to electrons gives too much "anti-gravitational" force for electrons. 
For now, we shall assume that there is some consistent 
framework that leads to the in-medium condensation of $H_{ij}$
along the lines of the proposals for the scalar field \cite{OP,KH}, although at this point it is an unproven assumption. 

Now we shall  consider interacting fields and answer the question of whether the variation of $\epsilon(z)$ 
could lead to the non-universality of clocks, or their abnormal speed-up or slow-down with depth, 
so that it could be picked up with dedicated experiments. (The connection between varying $c$ and coupling constants was 
previously discussed in Ref. \cite{BM}.)
For simplicity, let us consider the Lagrangian density of scalar quantum electrodynamics (QED) as a 
simplest model with gauge interactions, which we shall treat as a proxy to standard model. 
The unperturbed Lagrangian and its $\epsilon$-deformation are given by 
\begin{eqnarray}
{\cal L}_{\rm QED} = -\frac{1}{4}F_{\mu\nu}^2 +|(\partial_\mu  + igA_\mu )\Phi|^2 -m^2|\Phi|^2
\\ 
{\cal L}_\epsilon =h_{\mu\nu}\left(-F_{\mu\alpha}F_{\nu\alpha} +
2[(\partial_\mu  + igA_\mu )\Phi]^* (\partial_\nu +igA_\nu)\Phi \right)\\
-\frac13h_{\mu\mu}\left(\delta_1(|D_\mu\Phi|^2 -m^2|\Phi|^2)-\delta_2\frac{1}{4}F_{\mu\nu}^2  \right).
\nonumber
\end{eqnarray}
In these expressions, $g$ is the gauge coupling and $D_\mu = \partial_\mu + i g A_\mu $ is the covariant 
derivative, and trace is defined as $h_{\mu\mu}=h_{\mu\nu}\eta_{\mu\nu}$. The two free parameters $\delta_1$ and $\delta_2$ 
are meant to be order one, and they parametrize the model dependence. That is, in the first order in $\epsilon$ their values do not affect the modification of the propagation speed. We now go to the ansatz (\ref{epsilon}) and take
the adiabatic approximation where gradients of $\epsilon$ are neglected.
Then the sum of ${\cal L}_{\rm QED}$ and ${\cal L}_{\epsilon}$ gives
\begin{eqnarray}\nonumber
{\cal L}_{\rm QED+\epsilon} = {\cal L}_{\rm int}+(|\partial_0\Phi|^2-m^2|\Phi|^2)(1+\epsilon\delta_1) \\- |\partial_i \Phi|^2(1+\epsilon(2+\delta_1))~~~~~~~~~~\label{phigamma}\\
+\frac12{\bf E}^2(1+\epsilon(2-\delta_2)) -\frac12{\bf B}^2(1+\epsilon(4-\delta_2)),
\nonumber
\end{eqnarray}
where we separated terms bilinear in the fields from interactions. As evident from (\ref{phigamma}),
the propagation speeds of photons and charged scalars are the same, $c_{\gamma}=c_{\Phi}=
1+\epsilon$ up to $O(\epsilon^2)$ corrections, and independent on $\delta_{1(2)}$. 

In order to determine whether one should expect abnormal effects with clocks at $O(\epsilon)$ level, we 
make redefinitions of fields $\Phi$ and $A_\mu$ and distances $dx_i$, while leaving the time variable unchanged. Selecting $A_0=0$ gauge, we have:
\begin{eqnarray}
\Phi' = \Phi\left(1+\frac{\epsilon}{2}(3+\delta_1)\right);~ dx'=dx(1-\epsilon);\nonumber\\
 A_i'=A_i\left(1+\frac{\epsilon}{2}(5-\delta_2)\right).    ~~~~~~~~~~~~
\end{eqnarray}
Making these changes in the action, $S= \int d^4x{\cal L}$, and dropping primes over $x,~\Phi,~A_i$,
we read off a redefined equivalent Lagrangian to $O(\epsilon)$
level: 
\begin{eqnarray}
{\cal L}_{\rm QED+\epsilon} = -\frac{1}{4}F_{\mu\nu}^2   -m^2|\Phi|^2
\nonumber \\ 
+|(\partial_\mu+ig(1-\frac{\epsilon}{2}(3-\delta_2))A_\mu )\Phi|^2.
\label{lin}
\end{eqnarray}
Thus, we have the same scalar QED theory, but the coupling constant is now 
changed to 
\begin{equation}
\alpha_{\rm eff} = \left[   g(1-\frac{\epsilon}{2}(3-\delta_2))    \right]^2 = \alpha(1-\epsilon\times(3-\delta_2)),
\label{alpha_mod}
\end{equation}
and is a function of depth, following the $\epsilon(z)$ dependence. Notice that we do not have a change in mass of $\Phi$ as a consequences of us choosing couplings of $\epsilon$ to $|\partial_0 \Phi|^2-m^2|\Phi|^2$ combination. 

Different coupling constant means that the clocks
 build from "$\Phi$-matter" working on the atomic transition 
of $\Phi-\Phi^*$ bound state ($\sim \alpha^2m$) will see the abnormal $O(\epsilon)$ difference 
when placed below $z_0$. Also, different types of clocks with non-universal dependence on $\alpha$ will be 
sensitive to the $\epsilon$-induced change of frequencies. 
Elaborating on this, modification of couplings (\ref{alpha_mod}) implies two effects for the atomic clocks.
Imagine a hydrogen atom that serves as atomic clock that uses optical and hyperfine transitions, 
$ \hbar\omega_{\rm opt} = \frac{3}{8}\alpha^2 m_ec^2$ and 
$\hbar\omega_{\rm hf} = \frac{4g_pm_e}{3m_p}\alpha^4m_ec^2$, 
where we used the proton magnetic $g_p$ factor, and have restored $\hbar$ and $c$ for
clarity. Then comparison of the optical or hyperfines frequencies on the surface and underground imply the following 
ratio of frequencies,
\begin{eqnarray}
\label{effect1}
\frac{\omega_{\rm opt}(z<z_0)}{\omega_{\rm opt}(z=0)}= 1- 2\epsilon\times(3-\delta_2),\\
\frac{\omega_{\rm hf}(z<z_0)}{\omega_{\rm hf}(z=0)}= 1- 4\epsilon\times(3-\delta_2),
\end{eqnarray}
while the comparison of optical and hyperfine transition by clocks underground should give a modified 
ratio:
\begin{equation}
\label{effect2}
\frac{\omega_{\rm hf}}{\omega_{\rm opt}}(z<z_0) = \frac{32g_pm_e}{9m_p}\alpha^2(1- 2\epsilon\times(3-\delta_2)).
\end{equation} 
Using Eqs. (\ref{effect1})-(\ref{effect2}) as examples  it is easy to  predict what will happen for  arbitrary clocks comparison since alpha dependence has been
calculated for all known and proposed clocks -see {\em e.g.} Ref. \cite{Flambaum}.
Note that the dependence of the atomic unit of energy $\alpha^2 mc^2$  on $\epsilon$ (given in Eq. (\ref{effect1}) must be included when calculating dependence of the frequencies on $\epsilon$.
This dependence was not included in all the previous calculations of alpha dependence in Ref. \cite{Flambaum} since the results were presented in atomic units. For the purpose of the present
work this dependence can be easily restored by adding factor $\alpha^2$ to all frequencies calculated in \cite{Flambaum}. Then one should replace $\alpha$ by $\alpha_{\rm eff}$ from Eq. (\ref{alpha_mod}).

Two things are further worth noticing: firstly, the modification of the limiting 
propagation speed for two species, $A_\mu$ and $\Phi$,
does not carry an unambiguous prediction for $\alpha_{\rm eff}$, as it depends on the free parameter $\delta_2$ not 
fixed by the requirement of the universality of $c$. On account of that the naive logic "$c$ gets larger so that 
$\alpha = g^2/(\hbar c)$ gets smaller" does not hold. Secondly, there is a choice of $\delta_3=3$ when the 
${\cal L}_{\rm QED} \equiv {\cal L}_{\rm QED+\epsilon}$. This choice 
corresponds to a situation when $h_{\mu\nu}$ 
couples to the electromagnetic stress-energy tensor, 
$h_{\mu\nu}T_{\mu\nu}$, exactly as the linearized gravity would. 
In this case, to linear order in $\epsilon$, the clock universality will be preserved. 

Noting that $\epsilon$ suggested by the original  OPERA measurement (\ref{opera}) is very large relative to the precision of modern metrology, it is also interesting to investigate whether coupling of the background to the stress-energy operator would 
induce clock non-universality in $O(\epsilon^2)$ order. To that effect, we choose specific values of $\delta_{1(2)}$,
\begin{equation}
\label{Tmunu}
{\cal L}_{\rm QED+\epsilon} = {\cal L}_{\rm QED}+h_{\mu\nu}T^{\rm total}_{\mu\nu},
\end{equation}
and follow the similar procedure to find that to $O(\epsilon^2)$ order this theory is equivalent to 
\begin{eqnarray}
{\cal L}_{\rm QED+\epsilon} = \frac{1}{2}({\bf E}^2 -{\bf B}^2(1-4\epsilon^2))  -m^2|\Phi|^2
\nonumber \\ 
+|(\partial_\mu+ig\left(1-\frac{3\epsilon^2}{4}\right)A_\mu )\Phi|^2.
\label{quadr}
\end{eqnarray} 
Notice that the 
modification  (\ref{Tmunu}) creates $O(\epsilon^2)$ non-universality in the propagation speed of 
$A_\mu$ and $\Phi$, and this is why the $\epsilon^2$-dependence persists for the free fields. Even if one neglects 
magnetic effects, the coupling constant is modified at $O(\epsilon^2)$ level. Only the 
full general-relativity-(GR)-like extension $\eta_{\mu\nu} \to \eta_{\mu\nu} + h_{\mu\nu}$ of the original theory would preserve 
clock universality, which would entail additional $O(h_{\mu\nu}^2)$ terms in (\ref{Tmunu}) with 
specific coefficients. Thus, on the basis of (\ref{lin}) and (\ref{quadr}), we conclude that barring a very special GR-like case,
the modification of the limiting propagation speed for particles leads to the clock non-universality.

\section{Discussion: testing clocks at large depths} 

We have shown that one logical possibility - depth-dependent modification of $c$ for all species - 
is not immediately ruled out by the variety of constraints on LV and by gravity tests. Moreover, such possiblity can be tested with the neutrino speed experiments. With the expansion of the baseline it is 
conceivable to probe the propagation speed as deep as $O(100~{\rm km})$. 

The purpose of this 
paper was not to built an explicit dynamical model with {\em e.g.} condensation of spin-2 fields, but 
to investigate whether depth-induced modification of the maximum 
propagation speed can be seen with more conventional means other than timing of neutrino 
events. We have shown that with  a unique exception of pure GR-like coupling, one should expect an $O(\epsilon)$ (or $O(\epsilon^2)$ in case of $h_{\mu\nu}T^{\mu\nu}$ coupling) deviations of couplings 
from their "surface" values, spatially linked to the deviation of propagation speed. 
An assumption that buidling a self-consistent dynamical models
of this kind  is possible is of course a "leap of faith", and several arguments can be presented 
why this is difficult to achieve. 


Over the years, there has been a concerted effort to test the GR theory in space \cite{Will}.
Here we argue that the OPERA, MINOS and ICARUS results can be viewed as  first metrological tests
reaching 11 km underground. They give further insentives to 
test  clock universality and their abnormal speed-up/slow-down at great depths.
The GR-caused change in the frequency of clocks at depth $h$ and on the surface is given by 
$g_{\rm}h/c^2$ and for $h\sim 10~{\rm km}$ is $\Delta \omega/\omega \sim 10^{-12}$. Even the 
$\epsilon_{\rm OPERA}^2$-size effect is larger than GR shift by several orders of magnitude.  
Leaving aside the issue of the neutrino velocity, we believe that precision measurements underground are justified in their own right,
as no systematic tests of this kind were ever performed. 

One could envisage two types of experimental set-ups to test the constants-changing-with-depth
conjecture. First one could use the existing stable frequency emitters with known 
dependence on fundamental constants. Lowering them at great depths and comparing their
frequencies either in-situ or by transmitting the signals to the surface will test clock universality. 
The second set-up uses two identical clocks synchronized on 
the surface, with one of them brought deep underwater/underground for a period of time, with eventual 
comparison of time measured by both clocks upon the return. Should any of such experiments indeed detect larger-than-GR effects of the depth on clocks, once could also experimentally determine $z_0$, and further 
investigate possible large gradient effects around $z=z_0$. 

The deep underground locations could be used as a starting point for such tests. Indeed, deepest 
mines used for the underground science, such as {\em e.g.} Sudbury mine reach depths of 
2 km. Similar depths in ice are reached by the IceCube collaborations operating at the South Pole. 
The deepest comercialy used mines in South Africa extend 3.9 km underground. Ultimately, 
the "dream location" for such tests could be deep oceanic trenches and the deepest boreholes
that extend as deep as 11 km or more (which is incidentally very close to the maximum depth 
along the OPERA and ICARUS neutrino trajectory).

 The connection between variations of the fundamental constants and the variation of the
dimensionless ratios of the transition frequencies of different atomic clocks
is a well-researched subject. Current experimental sensitivity to the
 variation of $\alpha$ is better than  1 part in $10^{16}$ \cite{Hg+,Dy,Yb,Cs} 
per year, far exceeding accuracy needs discussed in this paper.
To check the link
 between the OPERA anomaly and variation of the fundamental constants it would
 be sufficient to compare commercially available $Cs$, $Rb$ or
 quartz clocks which have accuracy $10^{-11} - 10^{-12}$. The sensitivity of such
clocks to the variation of the fundamental constants has been calculated in 
\cite{Flambaum}. Further six orders of magnitude improvement in sensitivity may be
 reached using optical clocks and the frequency comb, as well as with the use of atomic systems 
with closely degenerate levels \cite{newref}.

Finally, we would like to comment that the density for many atomic clocks is very low, so that the 
question arises whether this would restore "surface values" for couplings inside atomic clocks deep underground. This question was answered in the models of chameleon-like varying constants \cite{OP}, 
where it was shown that the realistic model parameters ensure that the attainable sizes of the cavity are 
much less than the Compton wavelength of the chameleon field. In this case, the surface values
 for couplings will not be restored within the clock volume, if it is placed in an environment of non-zero $\epsilon$. 


The authors thank Drs. D. Budker, G,. Dvali and H. Mueller for useful discussions and communications. 
MP and VF would like to thank the  NZ IAS, Massey University, for the hospitality extended to them 
during their work on this project.


\begin{thebibliography}{10}

\bibitem{Opera}T.~Adam {\it et al.}  [OPERA Collaboration],
  arXiv:1109.4897 [hep-ex].

\bibitem{Minos} P.~Adamson {\it et al.}  [MINOS Collaboration],
  Phys.\ Rev.\  D {\bf 76}, 072005 (2007)
  [arXiv:0706.0437 [hep-ex]].

\bibitem{Icarus}  M.~Antonello {\it et al.}  [ICARUS Collaboration],
  arXiv:1203.3433 [hep-ex].

\bibitem{CG}  A.~G.~Cohen and S.~L.~Glashow,
  Phys.\ Rev.\ Lett.\  {\bf 107}, 181803 (2011)
  [arXiv:1109.6562 [hep-ph]].

\bibitem{nopiondecay}  L.~Gonzalez-Mestres,
  arXiv:1109.6630 [physics.gen-ph].

\bibitem{DV} G.~Dvali and A.~Vikman,
  arXiv:1109.5685 [hep-ph].

\bibitem{Brod} S.~J.~Brodsky and S.~Gardner,
  arXiv:1112.1090 [hep-ph].

\bibitem{MP} I.~Mocioiu and M.~Pospelov,
  Phys.\ Lett.\  B {\bf 534}, 114 (2002)
  [arXiv:hep-ph/0202160].

\bibitem{Sergei} G.~F.~Giudice, S.~Sibiryakov and A.~Strumia,
  arXiv:1109.5682 [hep-ph].




\bibitem{Heb}A.~Hebecker and A.~Knochel,
  arXiv:1111.6579 [hep-ph].

\bibitem{Brax} P.~Brax,
  arXiv:1202.0740 [hep-ph].

\bibitem{OP} K.~A.~Olive and M.~Pospelov,
  Phys.\ Rev.\  D {\bf 77}, 043524 (2008)
  [arXiv:0709.3825 [hep-ph]].

\bibitem{DN}T.~Damour and K.~Nordtvedt,
  Phys.\ Rev.\ Lett.\  {\bf 70}, 2217 (1993).

\bibitem{PD} T.~Damour and A.~M.~Polyakov,
  Nucl.\ Phys.\ B {\bf 423}, 532 (1994)
  [hep-th/9401069].

\bibitem{KW} J.~Khoury and A.~Weltman,
  Phys.\ Rev.\ Lett.\  {\bf 93}, 171104 (2004)
  [astro-ph/0309300].

\bibitem{Webb}   J.~K.~Webb, M.~T.~Murphy, V.~V.~Flambaum, V.~A.~Dzuba, J.~D.~Barrow, C.~W.~Churchill, J.~X.~Prochaska and A.~M.~Wolfe,
  Phys.\ Rev.\ Lett.\  {\bf 87}, 091301 (2001)
  [astro-ph/0012539].

\bibitem{KH}  K.~Hinterbichler and J.~Khoury,
  Phys.\ Rev.\ Lett.\  {\bf 104}, 231301 (2010)
  [arXiv:1001.4525 [hep-th]].

\bibitem{BM} J.~D.~Barrow and J.~Magueijo,
  Phys.\ Lett.\ B {\bf 443}, 104 (1998)
  [astro-ph/9811072].

\bibitem{Will}
 C.~M.~Will,
  Living Rev.\ Rel.\  {\bf 9}, 3 (2005)
  [gr-qc/0510072].


\bibitem{Hg+} T.~M.~Fortier {\em et al.}, Phys.\ Rev.\ Lett.\  {\bf 98}, 070801 (2007).

\bibitem{Dy}A.~Cingoz {\em et. al.},
  Phys.\ Rev.\ Lett.\  {\bf 98}, 040801 (2007)
  [physics/0609014].

\bibitem{Cs} S. Bize {\em et al.}, J. Phys. B: At. Mol. Opt. Phys. 38, S449 (2005).

\bibitem{Yb} E.~Peik {\em et al.},
  Phys.\ Rev.\ Lett.\  {\bf 93}, 170801 (2004).


\bibitem{Flambaum}
V.~V.~Flambaum, D.~B.~Leinweber, A.~W.~Thomas and R.~D.~Young,
  Phys.\ Rev.\ D {\bf 69}, 115006 (2004)
  [hep-ph/0402098];
  V.~V.~Flambaum and A.~F.~Tedesco,
  Phys.\ Rev.\ C {\bf 73}, 055501 (2006)
  [nucl-th/0601050];
T.~H.~Dinh, A.~Dunning, V.~A.~Dzuba and V.~V.~Flambaum,
  Phys.\ Rev.\ A {\bf 79}, 054102 (2009)
  [arXiv:0903.2090 [physics.atom-ph]]; M.~E.~Tobar, P.~Wolf, S.~Bize, G.~Santarelli and V.~Flambaum,
  Phys.\ Rev.\ D {\bf 81}, 022003 (2010)
  [arXiv:0912.2803 [gr-qc]];
J. C. Berengut, V. A. Dzuba, and V. V. Flambaum,   Phys. Rev. {\bf  A84}, 054501 (2011);
 E. J. Angstmann, V. A. Dzuba, and V. V. Flambaum. Phys. Rev. {\bf A70}, 014102, (2004);
 V. A. Dzuba, V. V. Flambaum, M. V. Marchenko.   Phys. Rev. {\bf A68}, 022506  (2003);
V.A. Dzuba, V.V. Flambaum,  Phys. Rev. {\bf A61}, 034502 (2000).

\bibitem{newref}
V. A. Dzuba,  V.V. Flambaum, J. K. Webb ,
 Phys. Rev. Lett. {\bf 82}, 888 ( 1999);
S. J. Ferrell {\em et al.},
 Phys. Rev.{\bf  A76}, 062104 (2007);
A. Cingoz {\em et al.}, Phys. Rev. Lett. {\bf 98}, 040801 (2007).


\end{thebibliography}
\end{document}